\documentclass[11pt,a4paper]{article}

\newcommand{\Z}{\mathbb{Z}}
\newcommand{\C}{\mathbb{C}}
\newcommand{\CP}{\mathbb{C}{\bf P}}
\newcommand{\M}{\mathcal{M}}
\newcommand{\R}{\mathbb{R}}
\newcommand{\Hom}{\mathrm{Hom}}
\newcommand{\Tr}{\mathrm{Tr}}
\newcommand{\id}{\mathbbm{1}}

\usepackage{bm,bbm,amsfonts,mathrsfs}
\usepackage{graphicx}
\usepackage{hyperref}

\usepackage{youngtab}

\makeatletter

\@addtoreset{equation}{section}
\makeatother

\date{\today}

\begin{document}

\begin{titlepage}

\renewcommand{\thefootnote}{\fnsymbol{footnote}}

\begin{flushright}
RIKEN-MP-26
\\
\end{flushright}

\vskip5em

\begin{center}
 {\LARGE {\bf 
Vortices on Orbifolds
 }}

 \vskip3em

 {\sc Taro Kimura}\footnote{E-mail address: 
 \href{mailto:kimura@dice.c.u-tokyo.ac.jp}
 {\tt kimura@dice.c.u-tokyo.ac.jp}}$^{1,2}$
 and
{\sc Muneto Nitta}\footnote{E-mail address:
\href{mailto:nitta@phys-h.keio.ac.jp}
 {\tt nitta@phys-h.keio.ac.jp}}$^3$

 \vskip2em

$^1${\it Department of Basic Science, University of Tokyo, 
 Tokyo 153-8902, Japan}\\ \vskip.2em
$^2${\it Mathematical Physics Lab., RIKEN Nishina Center, Saitama 351-0198,
 Japan 
}\\ \vskip.2em
$^3${\it Department of Physics, and Research and Education Center for
 Natural Sciences,\\ Keio University, Hiyoshi 4-4-1, Kanagawa 223-8521, Japan}

 \vskip3em

\end{center}

 \vskip2em

\begin{abstract}
The Abelian and non-Abelian vortices on orbifolds are investigated based
 on the moduli matrix approach, which is a powerful method to deal with
 the BPS equation.
The moduli space and the vortex collision are discussed through the
 moduli matrix as well as the regular space.
It is also shown that a quiver structure is found in the K\"ahler
 quotient, and {\em a half of ADHM} is obtained for the vortex theory
 on the orbifolds as the case before orbifolding.
\end{abstract}

\end{titlepage}

\tableofcontents

\setcounter{footnote}{0}

\section{Introduction}\label{sec:intro}

Vortices in Abelian gauge theory are essential degrees of freedom 
in superconductors under the magnetic fields \cite{Abrikosov:1956sx} 
while they also provide a model of 
strings in relativistic field theories \cite{Nielsen:1973cs}. 
They are refered as Abrikosov-Nielsen-Olesen (ANO) vortices.
In the case of the critical coupling 
between type I and II superconductors 
they saturate Bogomol'nyi bound and become 
Bogomol'nyi-Prasad-Sommerfield (BPS) states \cite{Bogomolny:1975de}. 
While in this case they can be embedded into supersymmetric theories 
and preserve a half of supersymmetry \cite{Witten:1978mh} on one hand, 
the whole solutions admit integration constants 
(moduli parameters, or collective coordinates) 
which constitute the moduli space on the other hand. 
A merit of the latter is that 
low-energy dynamics of vortices can be described by geodesics on 
the moduli space, see \cite{MantonSutcliffe200710} as a review.

Since the discovery of non-Abelian vortices
\cite{Hanany:2003hp,Auzzi:2003fs}, various aspects of vortices have been
investigated from both of string theory and gauge field theory perspectives.
One peculiar property is that a non-Abelian vortex in 
$U(N)$ gauge theory admits a $\CP^{N-1}$ moduli space 
which corresponds to Nambu-Goldstone modes 
in the internal space 
as for Yang-Mills instantons. 
Non-Abelian vortices provide a connection between 
BPS spectra of four-dimensional ${\cal N}=2$ supersymmetric gauge theories 
and two dimensional ${\cal N}=(2,2)$ 
$\CP^{N-1}$ model \cite{Shifman:2004dr,Hanany:2004ea}.
While 't Hooft-Polyakov monopoles become kinks 
inside a non-Abelian vortex \cite{Tong:2003pz},
Yang-Mills instantons become sigma model instantons there 
\cite{Hanany:2004ea,Eto:2004rz}.
As for the moduli space perspective, 
 by applying a useful treatment of BPS equations to the
non-Abelian vortex \cite{Eto:2005yh}, several properties 
on the moduli space and low-energy dynamics have been studied in detail \cite{Eto:2006pg,Eto:2006uw,Eto:2006cx,Eto:2006db,Fujimori:2010fk,
Eto:2011pj}.
Recently further developments, for example, vortex counting
\cite{Dimofte:2010tz,Yoshida:2011au,Bonelli:2011fq,Bonelli:2011wx},
volume of the moduli space \cite{Miyake:2011yr} and so on, have been
performed by applying the localization formula as well as the four
dimensional gauge theory.
They play an important role in
determining the low energy behavior of
the supersymmetric gauge theory.

Other than the flat space $\R^2 \simeq \C$,
studies of non-Abelian vortices have been so far
restricted to those on regular spaces,
such as a cylinder \cite{Eto:2006mz},
a torus \cite{Eto:2007aw,Lozano:2007bk},
Riemann surfaces
\cite{Popov:2008gw,Baptista:2008ex,Baptista:2010rv,Manton:2010mj},
and hyperbolic surfaces \cite{Manton:2010wu}.
(For Abelian vortices on various geometry,
see \cite{MantonSutcliffe200710} as a review,
after which there have been developments in
those on Riemann surfaces \cite{Popov:2007ms} and
hyperbolic surfaces \cite{Krusch:2009tn,Manton:2009ja}.
)

In this paper we consider Abelian and non-Abelian vortices
on two dimensional singular spaces,
namely the orbifolds $\C/\Z_n$.
Although the four dimensional gauge theory on the singular space, or the smooth
space given by resolving its singularity, 
which is called the asymptotically locally Euclidean (ALE) space,
has been investigated in detail, various studies on the vortex theory are
mainly based on the regular space as denoted above.
Indeed, in the case of the four dimensional theory, the ALE space has a
hyper-K\"ahler metric \cite{Kronheimer:1989zs}, so that almost the same
procedure to construct instantons on such a space as the usual Euclidean
space $\R^4$, or the four sphere $S^4$, can be performed
\cite{springerlink:10.1007/BF01233429,springerlink:10.1007/BF01444534}.
Furthermore the instanton counting and its matrix model description have
been discussed for the ALE space \cite{Fucito:2004ry,Kimura:2011zf}.
The recent development on the four dimensional theory, which
is called the AGT relation \cite{Alday:2009aq}, is also studied for this
case \cite{Belavin:2011pp,Nishioka:2011jk,Bonelli:2011jx,Belavin:2011tb,Bonelli:2011kv}.
Therefore it is natural to expect that study of vortices on orbifolds 
would give a novel perspective to the vortex theory.

To deal with the orbifold theory we first
apply the moduli matrix approach 
\cite{Isozumi:2004vg,Eto:2005yh,Eto:2006pg} 
rather than Hanany-Tong's method.
It is because, for the former one, we can easily see the space-time
structure of the vortex configuration and the dynamics of vortices
\cite{Eto:2006db,Eto:2011pj} while it is slightly difficult to
see it for the latter one, which is based on the dual configuration.
We first consider the fields on the usual regular space $\R^2 \simeq
\C$, namely the universal cover of the orbifold, and then take the
orbifold projection to obtain the $\Z_n$ invariant sector of the
configuration.
Because of the singular property of the space, we have to assign the
boundary condition, which breaks the original gauge symmetry.
This symmetry breaking affects the symmetry of the moduli space, and
leads to its decomposition.\footnote{
We remark that a similar but different
situation is found when the twisted mass term
is introduced \cite{Tong:2003pz}.
The gauge symmetry is dynamically broken due to the twisted mass, 
and then 
interesting topological excitations such as
confined monopoles \cite{Tong:2003pz} and 
vortices stretched between 
domain walls \cite{Isozumi:2004vg} can be observed.
}
We find that $k$-vortices on $\C/\Z_n$ with $k<n$ are fixed at 
the orbifold singularity, 
as Yang-Mills instantons 
\cite{springerlink:10.1007/BF01233429,springerlink:10.1007/BF01444534} 
and fractional D-branes \cite{Douglas:1996sw} on the orbifolds 
or their resolution to ALE spaces.
We call them fractional vortices borrowing the terminology 
of D-branes.\footnote{
Rather different types of fractional vortices appear 
in different settings such as gauge theories which 
flow in infrared to nonlinear sigma models with deformed 
\cite{Collie:2009iz,Eto:2009bz}
or singular 
\cite{Eto:2009bz,Eto:2008qw} target spaces.
} 
If we combine $n$ vortices properly they can be free from the singularity. 
We show that mirror images have the same internal moduli $\CP^{N-1}$ 
for $U(N)$ gauge theory. 
We also find that 
ANO vortices at the singularity do not give any moduli, 
while fractional non-Abelian vortices at the singularity
give internal moduli. 

We then discuss the vortex collision for the orbifold theory.
Since elements of the moduli matrix are directly related to smooth
coordinates of the moduli space, we can follow vortex dynamics along
geodesics on the moduli space. 
Especially when we concentrate on a short time behavior
at collision moment, the geodesics can be approximated 
by straight lines in smooth coordinates and therefore
we do not need the actual metric \cite{Eto:2006db}. 
In this paper we concentrate on the collision at the origin of the
complex plane, which corresponds to the singular point of the orbifold,
because other points are essentially the same as the usual regular space.
Due to the singularity of the orbifold, we can see unusual scattering
behavior in this case.

Finally we also comment on Hanany-Tong's approach, 
or the K\"ahler quotient for the orbifold theory.
Starting with the K\"ahler quotient for vortices 
on the flat space $\C$,
we implement the orbifold projection as well as the moduli matrix approach.
We can perform a quite similar procedure as the instantons on the ALE
space, and then obtain {\em a half of ADHM on the ALE spaces 
of the $A_{n-1}$ type}
as discussed in the regular case.

This paper is organized as follows.
In section \ref{sec:orb} we first discuss how to deal with the orbifold
theory through the boundary condition.
Section \ref{sec:modmat} is the main part of this paper, which is
devoted to the moduli matrix approach.
We investigate both of Abelian and non-Abelian vortices, and extract the
moduli space for such a vortex configuration.
We show how to implement the orbifold projection to the moduli matrix.
We consider the vortex collision in section \ref{sec:collsn}.
The vortex dynamics is obtained by studying the geodesics on the moduli space.
In section \ref{sec:quot} we also discuss the K\"ahler quotient for the
orbifold theory, according to Hanany-Tong's approach.
Stressing the similarity between vortices and instantons on the
orbifold, we propose a generic form of the moduli space of the vortices
on the orbifold.
We finally summarize the results and comment on some applications in
section \ref{sec:summary}.

\section{Orbifolding and boundary conditions}\label{sec:orb}

\begin{figure}[t]
 \begin{center}
  \includegraphics[width=15em]{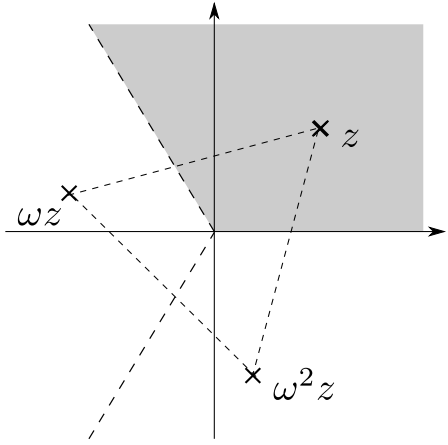}
 \end{center}
 \caption{Fundamental region of orbifold $\C/\Z_n$ for
 $n=3$. The orbifolding theory is obtained by identifying $z \sim \omega z$
 in the universal cover $\C$. $\omega=\exp(2\pi i/3)$ is the primitive
 third root of unity. The origin is the singular point.}
 \label{fig_orbifold}
\end{figure}

An orbifold $\C/\Z_n$ is constructed by identifying $z \sim \omega z$,
where 
$z$ is a coordinate of the covering space $\C$
and $\omega = \exp (2\pi i/n)$ is the primitive $n$-th root of unity, 
as shown in Fig.~\ref{fig_orbifold}.
For the Abelian case the following boundary condition 
on a Higgs scalar field $H(z,\bar{z})$
is allowed under this identification,
\begin{equation}
 H(z,\bar{z}) 
 \to H(\omega z, \overline{\omega z})
 = e^{i\alpha} H(z,\bar{z}) .
 \label{bc01}
\end{equation}
Due to the consistency condition 
for single valuedness of the Higgs field,
\begin{equation}
 H(z,\bar{z})
 = H(\omega^n z,\overline{\omega^n z}) 
 = e^{i n \alpha} H(z,\bar{z}) ,
\end{equation}
the phase factor $\alpha$ has to be quantized as
\begin{equation}
 \alpha = \frac{2 \pi}{n} m, \qquad
  m = 0, \cdots, n-1.
\end{equation}
Therefore the boundary condition (\ref{bc01}) is characterized by an
integer $0 \le m \le n-1$, and can be rewritten as
\begin{equation}
 H(\omega z, \overline{\omega z}) = \omega^m H(z,\bar{z}).
 \label{bc02}
\end{equation}
This phase factor suggests that the singularity of the spacetime assigns a
flux on the singular point via the boundary condition.

We then generalize this result to the non-Abelian theory. 
More precisely we introduce $U(N)$ gauge field 
and $N$ Higgs scalar fields in the fundamental representation, 
summarized as an $N \times N$ matrix $H(z, \overline{z})$ 
on which gauge symmetry acting from the left. 
In this case the Higgs and gauge field should be transformed as
\begin{equation}
 H(\omega z, \overline{\omega z}) = \Omega H(z,\bar{z}), \qquad
 A_z(\omega z, \overline{\omega z}) = \Omega A_z(z,\bar{z}) \Omega^{-1},
  \label{nonabelian_bc}
\end{equation}
where the orbifold transformation matrix $\Omega$, satisfying $\Omega^n = \id$,
can be diagonalized without loss of generality due to the gauge symmetry,
\begin{equation}
 \Omega = {\rm diag}(\omega^{m_1}, \cdots, \omega^{m_N}).
\end{equation}
A set of integers $(m_1, \cdots, m_N) \in \{0, \cdots, n-1\}^{N}$
characterizes the boundary condition of the Higgs field.
Changing the order of the diagonal component, it can be written as
\begin{equation}
 \Omega = 
  \left(
   \begin{array}{cccc}
    \omega^0 \id_{N^{(0)}} & & & 0 \\
    & \omega^1 \id_{N^{(1)}} & & \\
    & & \ddots & \\
    0 & & & \omega^{n-1} \id_{N^{(n-1)}} \\
   \end{array}
  \right). \label{eq:otm}
\end{equation}
This implies that (the global part of) 
the original gauge group $U(N)$ is broken by the boundary
condition as 
\begin{equation}
 U(N) \quad \longrightarrow 
  \quad U(N^{(0)}) \times \cdots \times U(N^{(n-1)})
\label{eq:gauge-breaking}
\end{equation}
where the rank of each gauge group is given by $N^{(m)} = \#\left\{ m_i
= m, i = 1, \cdots, N \right\}$, satisfying
\begin{equation}
 N^{(0)} + \cdots + N^{(n-1)} = N.
\end{equation}

\section{Moduli matrix approach}\label{sec:modmat}

The model we study in this paper is the following,
\begin{equation}
 \mathcal{L} = \Tr
  \left[
   - \frac{1}{2g^2} F_{\mu\nu} F^{\mu\nu}
   - \mathcal{D}_\mu H \mathcal{D}^\mu H^\dag
   - \lambda (c \id_N - HH^\dag)^2
  \right].
\end{equation}
Especially we focus on the critical coupling case $\lambda=g^2/4$.
In this case, the model can be endowed with supersymmetry 
by adding suitable bosonic fields and fermionic superpartners 
at least if we consider on the flat space. 
In supersymmetric context the parameter $c$ is called 
the Fayet-Illiopoulos parameter.
However supersymmetry is broken in orbifold theory \cite{Adams:2001sv} 
and so we do not discuss supersymmetry in this paper.
Note that BPS properties remain at least 
semi-clasically even without supersymmetry.

Then we can obtain the BPS equation from this Lagrangian,
\begin{equation}
 (D_1 + i D_2) H = 0, \qquad
 F_{12} + \frac{g^2}{2} ( c \id_N - HH^\dag) = 0.
\end{equation}
We then introduce the moduli matrix approach to solve this BPS equation.
The first equations can be solved as
\cite{Isozumi:2004vg,Eto:2005yh,Eto:2006pg} 
\begin{equation}
 H(z,\bar{z}) = S^{-1}(z,\bar{z}) H_0(z), \qquad
 A_z = A_1 + i A_2 = S^{-1}(z,\bar{z}) \partial_z S(z,\bar{z}), 
 \label{bc03}
\end{equation}
where the holomorphic matrix $H_0(z)$ is called the moduli matrix.
The rank of this matrix is $N$ for $U(N)$ gauge theory, thus it becomes
just a holomorphic function for the Abelian case.
Then the second equation can be recast into a gauge invariant 
equation, which has a unique solution 
\cite{Lin:2011zzf}.
Therefore all moduli parameters are contained in 
the moduli matrix $H_0(z)$ up to the following transformation;
This construction is invariant under
\begin{equation}
 \left(H_0(z),\ S(z,\bar{z}) \right) \longrightarrow
 \left(V(z) H_0(z),\ V(z) S(z,\bar{z}) \right)
\end{equation}
with $V(z) \in GL(N, \C)$ being holomorphic with respect to $z$.
This is called $V$-transformation and 
plays an important role on
classifying the moduli space.

Applying the boundary condition (\ref{nonabelian_bc}) the moduli matrix
behaves as
\begin{equation}
 H_0(\omega z) = \Omega H_0(z),
\end{equation}
since we have $\Omega H = \Omega S^{-1} (\Omega^{-1} \Omega) H_0$.
Especially the boundary condition for the moduli matrix plays an
important role on studying the moduli space of vortices on the orbifold.

Since the energy of this configuration is given by
\begin{equation}
 T = \frac{2 \pi c k }{n}
   = - i \frac{c}{2} \oint dz~\partial_z \log \det H_0(z),
\end{equation}
where the integral is performed only on the fundamental region, the
moduli matrix for $k$-vortex solution is written as
\begin{equation}
 \det H_0(z) = \prod_{i=1}^k (z - z_i) . \label{eq:detH_0}
\end{equation}
Here $k$ is the number of vortices on the whole complex plane $\C$ and
$z_i$ parametrizes the position of the vortex.
We remark the winding number $k/n$ is different from the vortex number $k$.
The former one can be fractional while the latter one is always integral.

\subsection{Abelian case}


We start with the simplest case $k=1$.
The moduli matrix, which is just a holomorphic function in this case,
can be generally written as
\begin{equation}
 H_0(z) = \prod_{i=1}^k (z - z_i) .
 \label{moduli_abelian}
\end{equation}
The parameter $z_i$ is regarded as a position of the $i$-th vortex.
For $k=1$, applying the boundary condition (\ref{bc03}), this function
has to satisfy
\begin{equation}
 \omega z - z_1 = \omega^m ( z - z_1).
\end{equation}
This condition is satisfied only if $m=1$ and $z_1 = 0$.
This means that the vortex is fixed at the singular point of $\C/\Z_n$, and
the boundary condition is automatically determined as $H_0(\omega z) =
\omega H_0(z)$.

We can easily generalize this result for $k < n$.
The boundary condition (\ref{bc02}) reduces the moduli matrix
(\ref{moduli_abelian}) to a trivial form
\begin{equation}
 H_{0;n}(z) = z^k ,
\end{equation}
satisfying $H_0(\omega z) = \omega^k H_0(z)$.
As the case of $k=1$, positions of vortices are fixed at the origin, and
thus the position moduli does not exist for $k < n$.
This means the fractional vortices are fixed at the origin.

We then study the case of $k=n$.
In this case the moduli matrix (\ref{moduli_abelian}) is reduced to
\begin{equation}
 H_{0;n}(z)
 = z^n - z_1^n
 = \prod_{p=0}^{n-1} (z - \omega^p z_1) , 
 \label{ANO_n}
\end{equation}
which satisfies $H_0(\omega z) = H_0(z)$.
This configuration has only one position modulus $z_1$
(see Fig.~\ref{fig_orbifold}).

Finally we provide the moduli matrix for generic vortex number $k = l
n + m \equiv m$~(mod~$n$),
\begin{equation}
 H_{0;n}(z)
 = z^m \prod_{i=1}^{l} (z^n - z_i^n)
 = z^m \prod_{i=1}^{l} \prod_{p=0}^{n-1} (z - \omega^p z_i).
\end{equation}
We can see that this solution satisfies the boundary condition $H_0(\omega z) =
\omega^m H_0(z)$.
Since there is only position, but internal moduli, the moduli space for
Abelian vortices on the orbifold is given by
\begin{equation}
 \M_{N=1,k;n} = \left(\C / \Z_n \right)^{\left[k/n\right]} /
  \mathfrak{S}_{\left[k/n\right]}.
\end{equation}
Here $[x]$ denotes the largest integer not greater than $x$.

We have found that $k$ fractional vortices with $k<n$ are 
fixed at the orbifold singularity of $\C/\Z_n$, 
but once $n$ vortices are combined together 
a set of them can be free from the singularity.
The same property can be found in 
Yang-Mills instantons 
\cite{springerlink:10.1007/BF01233429,springerlink:10.1007/BF01444534}
and fractional D-branes \cite{Douglas:1996sw} on orbifolds 
.


\subsection{$U(2)$ gauge theory}

\subsubsection{$k=1$ vortices}

We now consider the simplest non-Abelian gauge group $U(2)$ and 
the orbifold $\C/\Z_2$ for convenience.
The moduli matrix for $k=1$ vortex in $U(2)$ gauge theory on $\C$ is
given by \cite{Eto:2004rz,Eto:2006cx}
\begin{equation}
 H_0^{(1,0)}(z) = 
  \left(
   \begin{array}{cc}
    z - z_1 & 0 \\ -b' & 1
   \end{array}
  \right), \qquad
 H_0^{(0,1)}(z) = 
  \left(
   \begin{array}{cc}
    1 & -b \\ 0 & z - z_1
   \end{array}
  \right).
  \label{moduli_mat01}
\end{equation}
The corresponding orbifold transformation matrix can be simply determined by
reading the diagonal
components of the moduli matrix as
\begin{equation}
 \Omega^{(1,0)} = 
  \left(
   \begin{array}{cc}
    -1 & 0 \\ 0 & +1
   \end{array}
  \right), \qquad
 \Omega^{(0,1)} = 
  \left(
   \begin{array}{cc}
    +1 & 0 \\ 0 & -1
   \end{array}
  \right) .
\end{equation}
On the other hand, the moduli matrices (\ref{moduli_mat01}) themselves do
not satisfy the boundary condition in a consistent way because we have
different expressions as follows,
\begin{equation}
 H_0^{(1,0)}(-z) = 
  \left(
   \begin{array}{cc}
    - z - z_1 & 0 \\ -b' & 1
   \end{array}
  \right), \qquad
 \Omega^{(1,0)} H_0^{(1,0)}(z) = 
  \left(
   \begin{array}{cc}
    - z + z_1 & 0 \\ -b' & 1
   \end{array}
  \right) .
\end{equation}
By equating them, $H_0^{(1,0)}(-z)=\Omega^{(1,0)}H_0^{(1,0)}(z)$, 
we have to assign
$z_1 = 0$ to satisfy the boundary condition consistently.
We obtain the same condition from the other one $H_0^{(0,1)}(z)$.
Thus under this orbifold projection the moduli matrix for $\C/\Z_2$
theory becomes
\begin{equation}
 H_{0;n=2}^{(1,0)}(z) = 
  \left(
   \begin{array}{cc}
    z & 0 \\ -b' & 1
   \end{array}
  \right), \qquad
 H_{0,n=2}^{(0,1)}(z) = 
  \left(
   \begin{array}{cc}
    1 & -b \\ 0 & z
   \end{array}
  \right).
\end{equation}
We remark since $b$ and $b'$ are related as $b=1/b'$ via the
$V$-transformation, which is regarded as the complexified gauge
transformation \cite{Eto:2006cx}, there is only one parameter $b$
interpreted as the internal moduli. 
Therefore $k=1$ fractional vortex is fixed at the singular point, and
still has the internal degree of freedom, which turns out to be a coordinate of
$\CP^1$, as the usual non-Abelian vortex.
Indeed we can extract the orientation 
of the vortex via the following equations \cite{Eto:2006pg},
\begin{eqnarray}
  H_{0;n=2}^{(1,0)}(z=0) \vec\phi^{(1,0)} & = & 0, \\
  H_{0;n=2}^{(0,1)}(z=0) \vec\phi^{(0,1)} & = & 0.
\end{eqnarray}
Thus we have the orientational vectors
\begin{equation}
 \vec\phi^{(1,0)} = \left(\begin{array}{c} 1 \\ b' \end{array}\right), \qquad
 \vec\phi^{(0,1)} = \left(\begin{array}{c} b \\ 1 \end{array}\right).
\end{equation}
These are nothing but two patches of $\CP^1$.
Thus the moduli space for $k=1$ vortex solution in $U(2)$ gauge theory
on the orbifold $\C/\Z_2$ is given by
\begin{equation}
 \M_{N=2,k=1;n=2}^{\tiny\yng(1)} \simeq \{0\} \times \CP^1.
\end{equation}
Here we introduce a useful notation to characterize the moduli matrix by
using the Young diagram as discussed in \cite{Eto:2010aj}: $N$ entries
of the partition correspond to the powers of the diagonal components of
the moduli matrix for $U(N)$ gauge theory.
We can always obtain a descending ordered matrix by 
the $V$-transformation.
We again remark that there is no position moduli $\C/\Z_2$.
Unlike the Abelian case it still has internal moduli $\CP^1$, 
as in Yang-Mills instantons on orbifolds or their resolutions to 
ALE spaces 
\cite{springerlink:10.1007/BF01233429,springerlink:10.1007/BF01444534}.

\subsubsection{$k=2$ vortices}

Let us study a next example, configurations of $k=2$ vortices.
In this case generic forms of moduli matrices on $\C$ are given by
\begin{eqnarray}
 H_0^{(2,0)}(z) & = &
  \left(
   \begin{array}{cc}
    z^2 - \alpha' z - \beta' & 0 \\
    - a' z - b' & 1
   \end{array}
  \right), 
 \\ 
 H_0^{(1,1)}(z) & = & 
  \left(
   \begin{array}{cc}
    z - \phi & -\eta \\
    -\tilde \eta & z - \tilde \phi
   \end{array}
  \right), 
 \\ 
 H_0^{(0,2)}(z) & = & 
  \left(
   \begin{array}{cc}
    1 & -az-b \\
    0 & z^2 - \alpha z - \beta
   \end{array}
  \right).
\end{eqnarray}
The corresponding orbifold transformation matrices are given by
\begin{eqnarray}
 \Omega^{(2,0)} = \Omega^{(0,2)} =
  \left(
    \begin{array}{cc}
     +1 & 0 \\ 0 & +1
    \end{array}
  \right), \qquad
 \Omega^{(1,1)} =
  \left(
    \begin{array}{cc}
     -1 & 0 \\ 0 & -1
    \end{array}
  \right).
\end{eqnarray}
Performing the orbifold projection in a similar manner, the moduli
matrices are reduced to
\begin{eqnarray}
 H_{0;n=2}^{(2,0)}(z) & = &
  \left(
   \begin{array}{cc}
    z^2 - \beta' & 0 \\
     - b' & 1
   \end{array}
  \right) = H_0^{(1,0)}(z^2), 
  \label{moduli_mat_20} \\ 
 H_{0;n=2}^{(1,1)}(z) & = & 
  \left(
   \begin{array}{cc}
    z & 0 \\
    0 & z 
   \end{array}
  \right) 
  = z \id_2, 
  \label{moduli_mat_11} \\ 
 H_{0;n=2}^{(0,2)}(z) & = & 
  \left(
   \begin{array}{cc}
    1 & -b \\
    0 & z^2 - \beta
   \end{array}
  \right)
  = H_0^{(0,1)}(z^2) .
  \label{moduli_mat_02} 
\end{eqnarray}

We now comment on two remarkable facts.
First 
by imposing
\begin{eqnarray}
 \det H_{0;n=2}^{(2,0)}=\det H_{0;n=2}^{(0,2)} 
 \Rightarrow
 \beta = \beta'
\end{eqnarray}
in order for these two of the moduli matrices, $H_{0;n=2}^{(2,0)}$ 
and $H_{0;n=2}^{(0,2)}$, 
to describe the vortices in the same positions as can be seen in 
Eq.~(\ref{eq:detH_0}), 
they 
are almost the same as those for $k=1$ on $\C$ shown in
Eq.~(\ref{moduli_mat01}).
The difference is its argument, $z\to z^2$.
This means that two vortices located at 
$z=z_1=\sqrt \beta$ and $z=-z_1=-\sqrt \beta$ have the
same internal moduli $\CP^1$, and thus they turn out to be identical.
The vortex orientation can be obtained in a similar way as the case
$k=1$; 
from the equations 
\begin{eqnarray}
  H_{0;n=2}^{(2,0)}(z^2=\beta) \vec\phi^{(2,0)} & = & 0, \\
  H_{0;n=2}^{(0,2)}(z^2=\beta') \vec\phi^{(0,2)} & = & 0.
\end{eqnarray}
we obtain the orientational vectors 
\begin{equation}
 \vec\phi^{(2,0)} = \left(\begin{array}{c} 1 \\ b' \end{array}\right), \qquad
 \vec\phi^{(0,2)} = \left(\begin{array}{c} b \\ 1 \end{array}\right).
 \label{orientation_k2}
\end{equation}
They are nothing but coordinates of $\CP^1$.

Second is that we cannot perform a $V$-transformation which connects
$H_{0;n=2}^{(1,1)}$ with $H_{0;n=2}^{(2,0)}$ or $H_{0;n=2}^{(0,2)}$,
while $H_{0;n=2}^{(2,0)}$ and $H_{0;n=2}^{(0,2)}$ are connected via
$V$-transformation.
This is obvious from the fact that 
we cannot impose 
$\det H_{0;n=2}^{(1,1)} = \det H_{0;n=2}^{(2,0)} 
(= \det H_{0;n=2}^{(0,2)})$ except for $\beta=0$.
Therefore the part, corresponding to $H_{0;n=2}^{(1,1)}$, is
disconnected to the continuous one connecting $H_{0;n=2}^{(2,0)}$ and
$H_{0;n=2}^{(0,2)}$ in the moduli space $\M_{N=2,k=2;n=2}$.
In general,
 we can see if the orbifold transformation matrices are different, the
corresponding parts are disconnected.
Actually in this case we have
$\Omega^{(2,0)}=\Omega^{(0,2)}\not=\Omega^{(1,1)}$.


Furthermore the moduli matrix $H_{0;n=2}^{(1,1)}$ has neither of
position nor internal moduli, so that it corresponds to just an isolated
point in the moduli space.
The absence of internal moduli also means that 
it can be regarded as an Abelian (ANO) vortex. 
Actually we cannot extract the vortex orientation for this case because
$H_{0;n=2}^{(1,1)}(z=0) = 0$.
Thus implying $\Z_2$ transformation of a complex
coordinate $z \leftrightarrow -z$, the moduli space for $k=2$ turns out to be
\begin{equation}
  \M_{N=2,k=2;n=2} \simeq 
   \M_{N=2,k=2;n=2}^{\tiny\yng(2)} \cup
   \M_{N=2,k=2;n=2}^{\tiny\yng(1,1)},
\end{equation}
where each sector is given by
\begin{equation}
 \M_{N=2,k=2;n=2}^{\tiny\yng(2)} \simeq (\C/\Z_2) \times \CP^1, \qquad
 \M_{N=2,k=2;n=2}^{\tiny\yng(1,1)} \simeq \left\{0\right\} \times \{0\}.
\end{equation}

\subsubsection{$k=3$ vortices}

We then consider $k = 3$ vortices configurations.
The generic forms of the moduli matrices on $\C$ are written as
\begin{eqnarray}
 H_{0}^{(3,0)}(z) & = &
  \left(
   \begin{array}{cc}
    z^3 - \alpha' z^2 - \beta' z - \gamma' & 0 \\
    -a'z^2 - b' z - c' & 1
   \end{array}
  \right), \\
 H_{0}^{(2,1)}(z) & = &
  \left(
   \begin{array}{cc}
    z^2 - \eta' z - \kappa'  & 0  \\
    - \lambda' z - \xi' & z - \zeta'
   \end{array}
  \right), \\
 H_{0}^{(1,2)}(z) & = &
  \left(
   \begin{array}{cc}
    z - \zeta  & - \lambda z - \xi  \\
    0 & z^2 - \eta z - \kappa
   \end{array}
  \right), \\
 H_{0}^{(0,3)}(z) & = &
  \left(
   \begin{array}{cc}
    1 & -a z^2 - b z - c \\
    0 & z^3 - \alpha z^2 - \beta z - \gamma
   \end{array}
  \right).
\end{eqnarray}
The orbifold transformation matrices, characterizing the boundary conditions, can
be simply obtained from powers of the diagonal components of the moduli
matrices,
\begin{equation}
 \Omega^{(3,0)} = \Omega^{(1,2)}
  = \left(
     \begin{array}{cc}
      -1 & 0 \\ 0 & +1
     \end{array}
    \right), \qquad
 \Omega^{(2,1)} = \Omega^{(0,3)}
  = \left(
     \begin{array}{cc}
      +1 & 0 \\ 0 & -1
     \end{array}
    \right).
\end{equation}
We then apply the orbifold projection to the moduli matrices by
removing components whose powers are different by modulo $n$ from that of
the diagonal component in the same row.
The moduli matrices for the orbifolding theory are 
obtained as
\begin{eqnarray}
 H_{0;n=2}^{(3,0)}(z) & = &
  \left(
   \begin{array}{cc}
    z(z^2-\beta') & 0 \\ -a' z^2 - c' & 1
   \end{array}
  \right), \\
 H_{0;n=2}^{(2,1)}(z) & = &
  \left(
   \begin{array}{cc}
    z^2 - \kappa' & 0 \\ -\lambda' z & z
   \end{array}
  \right), \\
 H_{0;n=2}^{(1,2)}(z) & = &
  \left(
   \begin{array}{cc}
    z & -\lambda z \\
    0 & z^2 - \kappa
   \end{array}
  \right), \\
 H_{0;n=2}^{(0,3)}(z) & = &
 \left(
  \begin{array}{cc}
   1 & -a z^2 - c \\ 0 & z(z^2-\beta)
  \end{array}
 \right) .
\end{eqnarray}
As in the previous example 
we have to choose $\beta=\beta'$ 
and $\kappa=\kappa'$ 
to obtain the same determinant
$\det H_{0:n=2}^{(3,0)}=\det H_{0:n=2}^{(0,3)}$ 
and $\det H_{0:n=2}^{(2,1)}=\det H_{0:n=2}^{(1,2)}$, 
respectively. 
We can see that $H_{0;n=2}^{(3,0)}$ and $H_{0;n=2}^{(0,3)}$ or 
$H_{0;n=2}^{(2,1)}$ and $H_{0;n=2}^{(1,2)}$ are connected via 
$V$-transformations.
However we cannot connect a different combination, 
for example, $H_{0;n=2}^{(3,0)}$ and $H_{0;n=2}^{(2,1)}$, 
and they are disconnected as in the $k=2$ case.
Indeed the numbers of internal moduli are different: two for
the former and one for the latter.
Furthermore, the orbifold transformation matrices for them are different by
comparing in the {\it descending} order.
Although
we have the same orbifold transformation matrices for $H_{0;n=2}^{(3,0)}$
and $H_{0;n=2}^{(1,2)}$, 
we have to compare $\Omega^{(3,0)}$ with $\Omega^{(2,1)}$ 
because $\Omega^{(1,2)}$ is not in the {\it descending} order.

The vortex orientations are given by
\begin{equation}
 \vec\phi^{(3,0)} = \left(\begin{array}{c} 1 \\ a'\beta'+c' \end{array} \right)
 \quad \mbox{for} \quad z^2 = \beta, \qquad
 \vec\phi^{(3,0)} = \left(\begin{array}{c} 1 \\ c' \end{array} \right)
 \quad \mbox{for} \quad z = 0,
  \label{orientation_k3_1}
\end{equation}
\begin{equation}
 \vec\phi^{(0,3)} = \left(\begin{array}{c} a\beta+c\\ 1 \end{array} \right)
 \quad \mbox{for} \quad z^2 = \beta, \qquad
 \vec\phi^{(0,3)} = \left(\begin{array}{c} c \\ 1 \end{array} \right)
 \quad \mbox{for} \quad z = 0.
  \label{orientation_k3_2}
\end{equation}

The two internal moduli for $H_{0;n=2}^{(3,0)}$ and $H_{0;n=2}^{(0,3)}$
are interpreted as coordinates of $(\CP^1)^2$ for the separated $k=2$
vortices as well as the usual $k=2$ configuration on $\C$ 
studied in\cite{Eto:2006cx}.
When these $k=2$ vortices coincide, we have $\beta=0$.
In this case the remaining orientation is just a $\CP^1$.
This situation is essentially different from the the coincident $k=2$
vortices on $\C$ 
where the moduli space is given by 
$W\CP_{(2,1,1)}^2 \simeq \CP^2/\Z_2$ 
\cite{Auzzi:2005gr,Eto:2006cx}.

On the other hand, since there is the only one internal moduli for
$H_{0;n=2}^{(2,1)}$ and $H_{0;n=2}^{(1,2)}$, we have to discuss which
vortex possesses it.
Then taking an asymptotic limit $z \sim \pm \sqrt{\kappa} \to \infty$, we have
\begin{equation}
 H_{0;n=2}^{(2,1)}(z) 
  \longrightarrow
  \pm \sqrt{\kappa} \left(
       \begin{array}{cc}
	2(z \mp \sqrt{\kappa}) & 0 \\ -\lambda' & 1
       \end{array}
      \right),
\end{equation}
\begin{equation}
 H_{0;n=2}^{(1,2)}(z) 
  \longrightarrow
  \pm \sqrt{\kappa} \left(
       \begin{array}{cc}
	1 & -\lambda \\ 0 & 2(z \mp \sqrt{\kappa})
       \end{array}
      \right).
\end{equation}
This means that an internal moduli parameter $\lambda=1/\lambda'$,
regarded as a coordinate of $\CP^1$, belongs to a vortex at
$\pm\sqrt{\kappa}$.
Indeed the vortex orientations at $z^2 = \kappa$ turn out to be
\begin{equation}
 \vec\phi^{(2,1)} = \left(\begin{array}{c} 1 \\ \lambda' \end{array} \right),
 \qquad
 \vec\phi^{(1,2)} = \left(\begin{array}{c} \lambda \\ 1 \end{array} \right).
\end{equation}

In summary, in this case the moduli space is given by
\begin{equation}
 \M_{N=2, k=3, n=2} \simeq
  \M_{N=2, k=3, n=2}^{\tiny\yng(3)} \cup
  \M_{N=2, k=3, n=2}^{\tiny\yng(2,1)}.
\end{equation}
Here, for the first sector we have to write it as follows,
\begin{equation}
 \M_{N=2,k=3,n=2}^{\tiny\yng(3)} \simeq 
  \M_{\mathrm{separate}}^{\tiny\yng(3)} \cup
  \M_{\mathrm{coincident}}^{\tiny\yng(3)},
\end{equation}
where each part
\begin{equation}
 \M_{\mathrm{separate}}^{\tiny\yng(3)} 
  \simeq (\C^*/\Z_2) \times (\CP^1)^2, \qquad
 \M_{\mathrm{coincident}}^{\tiny\yng(3)} 
  \simeq 
  \{0\} \times \CP^1  
\end{equation}
is glued to each other.
Since the first sector is defined only when $\beta\not=0$, we denote the
position moduli as $\C^*=\C\backslash \{0\}$.
The second sector is simply given by
\begin{equation}
 \M_{N=2, k=3, n=2}^{\tiny\yng(2,1)} \simeq (\C/\Z_2) \times \CP^1.
\end{equation}

\subsection{$U(N)$ gauge theory}

Let us consider the moduli matrix for the generic $U(N)$ gauge theory.
Decomposing the total vortex number, $k \to (k_1,\cdots, k_N)$ with
$k_1+\cdots+k_N=k$ and $k_1 \ge k_2 \ge \cdots \ge k_N$, it is written
as a lower triangle matrix,
\begin{equation}
 H_0(z) =
  \left(
   \begin{array}{cccc}
    z^{k_1} + \cdots & & & 0 \\    
    \ast & z^{k_2} + \cdots & & \\
    \vdots & \ddots & \ddots & \\
    \ast & \cdots & \ast & z^{k_N} + \cdots \\
   \end{array}
  \right).
  \label{moduli_mat_N01}
\end{equation}
We can perform the orbifold projection as discussed in the previous section.
Writing each decomposed vortex number as $k_i = l_i n + m_i \equiv m_i$
(mod $n$), the orbifold transformation matrix is given by
\begin{equation}
 \Omega =
  \left(
   \begin{array}{cccc}
    \omega^{m_1} & & & 0 \\
    & \omega^{m_2} & & \\
    & & \ddots & \\
    0 & & & \omega^{m_N} \\
   \end{array}
  \right) .
  \label{trans_mat_N01}
\end{equation}
In order for the moduli matrix to
satisfy the boundary condition consistently, we
have to remove factors whose powers are different by modulo $n$ from that of
the diagonal component in the same row.
Thus the moduli matrix (\ref{moduli_mat_N01}) becomes
\begin{equation}
 H_{0;n}(z) =
  \left(
   \begin{array}{cccc}
    z^{m_1} \left(Z^{l_1} + \cdots \right) & & & 0 \\
    z^{m_2} f_{1,2}(Z) & z^{m_2} \left(Z^{l_2} + \cdots \right) & & \\
     \vdots & \ddots & \ddots & \\
    z^{m_N} f_{1,N}(Z) & \cdots & z^{m_N} f_{N-1,N}(Z) & z^{m_N}
     \left(Z^{l_N} + \cdots \right) \\
   \end{array}
  \right)
\end{equation}
where $f_{i,j}(Z)$ is a polynomial of $Z = z^n$ whose degree is lower
than $l_i$.
Its determinant is given by
\begin{equation}
 \det H_{0;n}(z) 
  = z^{m_1+\cdots+m_N} \prod_{i=1}^{l_1 + \cdots + l_N}
  \left( Z - Z_i \right)
  = z^{m_1+\cdots+m_N} \prod_{i=1}^{l_1 + \cdots + l_N}
  \left( z^n - z_i^n \right).
\end{equation}
This shows that the number of the position moduli is $l_1 + \cdots + l_N$,
and there are $m_1 + \cdots + m_N$ vortices fixed at the origin.
In terms of broken gauge group the number of fixed vortices is given by
\begin{equation}
 \sum_{i=1}^N m_i = \sum_{m=0}^{n-1} m N^{(m)} \le (n-1) N.
\end{equation}
This inequality is saturated when $N^{(0)}=\cdots=N^{(n-2)}=0$, $N^{(n-1)}=N$.

\section{Vortex collision}\label{sec:collsn}

Vortex collision is an important aspect of vortex dynamics.
The merit for the study with the moduli matrix is that it is directly related 
to coordinates of the moduli space.
It has been investigated by studying geodesics in the moduli space 
for vortices on $\C$.
Using a general formula for the moduli space metric \cite{Eto:2006uw} 
head-on collision during short time 
has been studied \cite{Eto:2006db},   
and asymptotic dynamics have been studied \cite{Eto:2011pj}
by using the asymptotic metric for well-separated vortices 
\cite{Fujimori:2010fk}. 
Here we concentrate on collision dynamics on the orbifold singularity 
by applying the method in \cite{Eto:2006db}.

\begin{figure}[t]
 \begin{center}
  \includegraphics[width=30em]{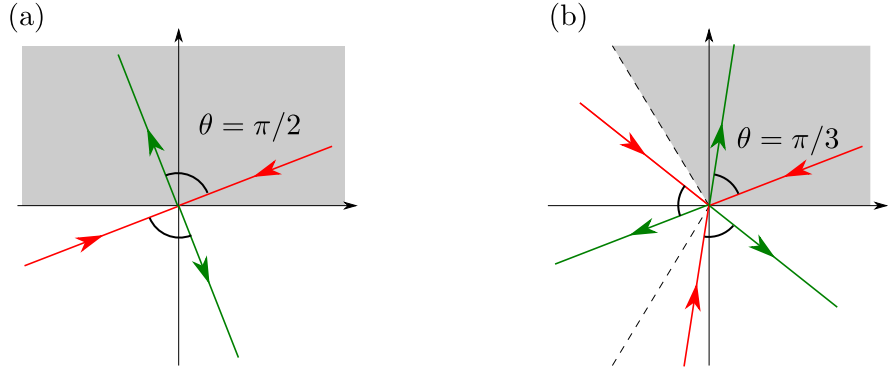}
 \end{center}
 \caption{Collision of vortices on the orbifold (a) $\C/\Z_2$ with
 $\theta=\pi/3$ and (b) $\C/\Z_3$ with $\theta = \pi/3$. 
 For $\Z_n$ with an odd $n$, vortices just look passing through the
 origin of the universal covering space $\C$.
 }
 \label{fig_collision}
\end{figure}

Let us start with $k=n$ Abelian vortices on the orbifold $\C/\Z_n$.
We now rewrite the moduli matrix, just a holomorphic function, given by
(\ref{ANO_n}) as
\begin{equation}
 H_{0;n}(z,t) = Z - \Xi t = \prod_{m=0}^{n-1} (z - \omega^m \xi t^{1/n})
\end{equation}
where $Z=z^n$, $\Xi = \xi^n$ and $t \in \R$.
After changing $t \to -t$ we have
\begin{equation}
 H_{0;n}(z,-t) = Z + \Xi t 
  = \prod_{m=0}^{n-1}(z - \omega^m e^{i \pi/n}  \xi t^{1/n}).
\end{equation}
Here we have an extra factor $e^{i\pi/n}$.
This means that vortices are colliding 
at $t=0$ 
at the origin with a scattering angle
$\theta = \pi/n$.%
\footnote{Such $\Z_n$ symmetric collisions are also studied
on hyperbolic surfaces \cite{Krusch:2009tn}.}
Fig.~\ref{fig_collision} shows collision of Abelian vortices on the
orbifolds $\C/\Z_2$ and $\C/\Z_3$. 
The moduli matrix approach correctly reproduces the results in \cite{Arthur:1995eh,MacKenzie:1995jw,MantonSutcliffe200710} 
in which $\Z_n$ symmetric collisions are studied in $\C$.

Let us then discuss the non-Abelian vortex collision.
This scattering property is also found for the non-Abelian cases
because it is only related to the position moduli.
On the other hand, we can see an interesting behavior of the internal
moduli.

For the case of $k=2$, $n=2$, the vortex orientation is given by
(\ref{orientation_k2}).
Although we expand the moduli matrices, (\ref{moduli_mat_20}) and
(\ref{moduli_mat_02}), as $\beta=\mathfrak{b}t$ and
$\beta'=\mathfrak{b}'t$, the vortex orientation is not affected at all.
This means that even though each collides and scatters 
with its mirror image with an angle $\theta=\pi/n=\pi/2$ in space, 
but its internal moduli do not change.

For the $k=3$ case we substitute $\beta=\mathfrak{b} t$ and
$\beta'=\mathfrak{b}'t$ as before.
The orientations, (\ref{orientation_k3_1}) and
(\ref{orientation_k3_2}), are expanded as
\begin{equation}
 \vec\phi^{(3,0)} = \left(\begin{array}{c} 1 \\ a'\mathfrak{b}'t+c'
			  \end{array} \right) \quad \mbox{for} \quad z^2
 = \mathfrak{b}'t, \qquad
 \vec\phi^{(3,0)} = \left(\begin{array}{c} 1 \\ c' \end{array} \right)
 \quad \mbox{for} \quad z = 0,
\end{equation}
\begin{equation}
 \vec\phi^{(0,3)} = \left(\begin{array}{c} a\mathfrak{b}t+c\\ 1
			  \end{array} \right) \quad \mbox{for} \quad z^2
 = \mathfrak{b}t, \qquad
 \vec\phi^{(0,3)} = \left(\begin{array}{c} c \\ 1 \end{array} \right)
 \quad \mbox{for} \quad z = 0.
\end{equation}
We can see that they coincide at $t=0$ as
\begin{equation}
 \vec\phi^{(3,0)} = \left(\begin{array}{c} 1 \\ c' \end{array}\right),
 \qquad
 \vec\phi^{(0,3)} = \left(\begin{array}{c} c \\ 1 \end{array}\right), 
\end{equation}
and go through the original direction even after the collision.
This behavior is different from the usual vortex collision on $\C$
\cite{Eto:2006db}.
In that case, after orientations of two colliding vortices coincide, 
they change their directions in the internal space.
On the other hand, in this case, because they have only one parameter,
$\mathfrak{b}$ or $\mathfrak{b}'$, the direction of the time-evolution
of the internal moduli does not change.
We remark that when we consider the $k=4$ configuration on $\C/\Z_2$, 
we will see the same situation as the usual $k=2$ vortex collision on $\C$.

\section{K\"ahler quotient}\label{sec:quot}

We then discuss the K\"ahler quotient description of the moduli space,
which has been originally studied in terms of string theory 
\cite{Hanany:2003hp} 
and later proven from field theory 
\cite{Eto:2005yh,Eto:2006pg}.
It is obtained from the D-term condition for the effective theory on
D-branes,
\begin{equation}
 [B,B^\dag] + I I^\dag = c \id_k.
  \label{Dterm01}
\end{equation}
Here we have $B \in \Hom(V,V)$, $I \in \Hom(V,W)$ for two vector spaces
$V$ and $W$.
The winding number and the rank of the gauge group are given by their
dimensions, dim~$V=k$ and dim~$W=N$.
Since we have $U(k)$ gauge symmetry for these data, $(B,I) \to
(gBg^{-1},gI)$, $g\in U(k)$, the moduli space is given by 
\begin{equation}
 \M_{N,k} \simeq \{(B,I)| [B,B^\dag] + I I^\dag = c \id_k \} / U(k).
\end{equation}

Let us consider the moduli space of vortices on orbifolds with respect
to this K\"ahler quotient description.
We introduce the decomposed vector spaces in order to characterize the
representations under $\Z_n$ action,
\begin{equation}
 V = \bigoplus_{m=0}^{n-1} V_m, \qquad
 W = \bigoplus_{m=0}^{n-1} W_m.
 \label{decom_vec_sp}
\end{equation}
Their dimensions are dim~$V_m = k^{(m)}$, dim~$W_m = N^{(m)}$ 
with $\sum_{m=0}^{n-1} k^{(m)} =k$ and 
$\sum_{m=0}^{n-1} N^{(m)} =N$.
This decomposition corresponds to the gauge symmetry breaking 
in Eq.~(\ref{eq:gauge-breaking})
due to the boundary condition as discussed in the previous section.

 \begin{figure}[t]
 \begin{center}
  \includegraphics[width=25em]{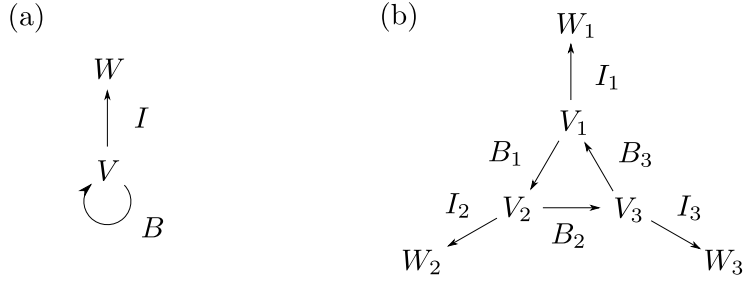}
 \end{center}
 \caption{Quiver diagrams for the moduli spaces of vortices on the
 orbifolds: (a) $\C$ and (b) $\C/\Z_3$. }
 \label{quiv_fig}
\end{figure}

The isometry $z \to e^{\epsilon} z$ of $\C$ acts on the data as $(B,I)
\to (e^\epsilon B, I)$.
Therefore we have to consider components of $B_m \in \Hom(V_m, V_{m+1})$
and $I_m \in \Hom(V_m, W_m)$ where we identify $V_n = V_0$ and $W_n =
W_0$.
Fig.~\ref{quiv_fig} shows quiver diagrams for vector spaces.
Comparing with the ADHM method for the ALE spaces, as shown in Appendix
\ref{sec:ADHM}, we can see the quiver for the orbifold vortex theory is
{\em a half of ADHM} as well as the usual case on $\C$.
We remark that, if the two dimensional orbifold $\C/\Gamma$ is considered, 
we have to set $\Gamma$ as a finite subgroup of $U(1)$.
This means it must be the cyclic group $\Gamma = \Z_n$,
corresponding to the $A_{n-1}$ type root system.

The algebraic condition (\ref{Dterm01}) yields
\begin{equation}
 B_{m-1} B_{m-1}^\dag - B_{m}^\dag B_{m} + I_m I_m^\dag = c \id_{k^{(m)}},
  \qquad
  m = 0, \cdots, n-1.
\end{equation}
We have $U(k^{(m)})$ symmetry for $m$-th equation.
Note that as well as the ALE spaces, if we could resolve
the singularity of the orbifold $\C/\Z_n$, 
a different FI parameter $c_m$ would be applied to each block.
However such a resolution does not exist for one dimensional 
orbifolds $\C/\Z_n$.
Thus the total moduli space is given by 
\begin{equation}
 \M_{\vec{N},\vec{k};n} \simeq 
  \frac{\left\{
   (B_0,\cdots,B_{n-1},I_0, \cdots, I_{n-1})| B_{m-1} B_{m-1}^\dag -
   B_{m}^\dag B_{m} + I_m I_m^\dag = c \id_{k^{(m)}}, 
   m=0, \cdots, n-1
  \right\}}{U(k^{(0)}) \times \cdots \times U(k^{(n-1)})} .
\end{equation}
This moduli space is not determined by just choosing the rank of the original
gauge group and the total winding number of vortices.
We have to specify a way of decomposition, thus it is labeled by
$\vec{N}=(N^{(0)}, \cdots, N^{(n-1)})$ and $\vec{k} = (k^{(0)}, \cdots,
k^{(n-1)})$, satisfying $N^{(0)}+\cdots+N^{(n-1)}=N$ and
$k^{(0)}+\cdots+k^{(n-1)}=k$.
Since this decomposition corresponds to the boundary condition as
discussed above, the moduli spaces under different decompositions
are disconnected.
The moduli space of $k$-vortex on the orbifold is given by
\begin{equation}
 \M_{N,k;n} = \bigcup_{|\vec{N}|=N, |\vec{k}|=k} 
  \M_{\vec{N},\vec{k};n}.
\end{equation}

\section{Summary and Discussion}\label{sec:summary}

We have investigated several properties of vortices on orbifolds 
$\C/\Z_n$.
First we have considered consistent boundary conditions for fields on 
the orbifolds.
Then we have performed the moduli matrix approach to characterize the moduli 
space of vortices.
Under the orbifold boundary condition the moduli matrices should be 
determined consistently.
We have clarified some restrictions to obtain the moduli matrices for 
the orbifolds from the usual ones for $\C$, which we call the orbifold 
projection.

We have investigated the moduli spaces based on the orbifolded moduli 
matrices with some examples.
The most remarkable point which we can find in the both of Abelian and 
non-Abelian cases is that the fractional vortices are just fixed 
at the singular point of the orbifold, namely the origin of the complex 
plane. 
They lose position moduli but fractional non-Abelian vortices 
still possess internal moduli 
$\CP^{N-1}$ for $U(N)$ gauge theory.
Indeed a similar situation for a fractional object is observed in other
models; Yang-Mills instantons
\cite{springerlink:10.1007/BF01233429,springerlink:10.1007/BF01444534}
and fractional D-branes \cite{Douglas:1996sw} on orbifolds or their
resolution to ALE spaces.

In the non-Abelian theory we have the internal moduli as well as the
usual vortices on $\C$.
However, while the whole moduli space is connected via the $V$-transformation
before orbifolding, the moduli space for the orbifold theory can be
decomposed to disconnected parts.
This decomposition is due to the gauge symmetry breaking via the 
boundary condition. 
We have found that the decomposed sectors 
one to one correspond to 
the orbifold transfomation matrices 
$\Omega$ in Eq.~(\ref{eq:otm}), 
which transform fields at $z$ and $\omega z$,
in the {\it descending} order. 
Although it is similar to the theory in presence of the twisted 
mass term \cite{Tong:2003pz}, 
an essential difference exists between them;  
each sector is completely decoupled from each other 
in the case of orbifolds while it is just 
separated by a potential in the case of twisted mass. 
Accordingly, two vortex states can be connected along 
a string making a confined monopole in the case of 
the twisted masses while it is impossible for the orbifolds.

We have also discussed vortex collision for the orbifold theory with the
moduli matrix perspective.
We have shown the scattering angle is directly reflecting the orbifold
structure.
Furthermore we can observe an unusual behavior of the internal moduli
for the non-Abelian vortex.
The internal orientations of two vortices coincide when they are
colliding, but their behavior after the collision is different from the
usual one on $\C$.

Finally the relation to Hanany-Tong's approach is discussed.
As well as the instantons on the orbifold, or the ALE space given by
resolving its singularity, we have a similar quiver theory for the K\"ahler
quotient.
We have seen that it is just a half of ADHM for
instantons on the type-$A$ orbifold $\C^2/\Z_n$.

Here before closing the paper, 
we give several discussions.

In this paper we have studied short time dynamics 
around the collision moment for which 
we have not needed explicit metric on the moduli space, 
but we need it for studying a long time behavior. 
The moduli space metric has been explicitly obtained 
for non-Abelian vortices on $\C$ \cite{Fujimori:2010fk}
and on Riemann surfaces \cite{Baptista:2010rv}
when vortices are well separated, 
and dynamics and scattering have been studied on $\C$ \cite{Eto:2011pj}.
Beyond our work 
it will be an interesting and important future problem to construct 
an explicit metric of the moduli space 
in our case of the orbifold $\C/\Z_n$. 
One nontrivial effect will be the existence of 
the fractional vortices fixed at the orbifold singularity. 
It does not appear in the moduli space in some cases, 
but even in that case the existence of such 
a flux at the singularity will change 
the moduli space metric. 
For instance, ANO vortices fixed at the singularity 
in both Abelian and non-Abelian gauge theories 
have no moduli but the fluxes do exist there.

In this paper we have concentrated on local vortices, 
namely vortices in $U(N)$ gauge theory with $N_{\rm F}=N$ 
fundamental scalar fields.
When the number of flavor is greater than the number of color, 
$N_{\rm F}>N$, vortices are called semi-local \cite{Vachaspati:1991dz}.  
However we have to be careful on (non-)normalizability of 
moduli parameters in this case \cite{Shifman:2006kd}.
Other extensions would be changing gauge symmetry 
from $U(N)$ to $G \times U(1)$ 
such as $G=SO,USp$ \cite{Eto:2008yi}.

Let us make a comment on a relation to Yang-Mills instantons. 
In five dimensional gauge theory in the Higgs phase, 
instantons on $\C^2$ can live stably in 
a vortex world-volume  
in which Yang-Mills instantons are regarded 
as lumps or sigma model instantons \cite{Eto:2004rz}.
It becomes Amoeba in more general on $(\C^*)^2$ \cite{Fujimori:2008ee}.
In our case of the orbifold theory, 
instantons live on a vortex in $\C \times \C/\Z_n$ 
where the vortex world-volume extends to $\C$.
In the case of $\C^2$ we obtain usual instantons 
in the limit of vanishing Fayet-Iliopoulos parameter $c$
in which the model goes to unbroken phase. 
It will be interesting to study what happens 
in the same limit for instantons on $\C \times \C/\Z_n$. 
Also this may be related to surface operators 
in the AGT relation \cite{Kanno:2011fw}.

Finally
we comment on physical applications of the vortices on orbifolds.
The vortex polygon and crystal, discussed in a few body vortex system
\cite{dhanak1992,Aref20031}, have a similar property to the orbifold
theory studied in this paper.
We have to discuss relation between them in a future work.
Another proposal is application to the two dimensional carbon system.
The conical structure, mimicking the orbifolds $\C/\Z_n$, could be
obtained by manipulating the graphene sheet or the carbon nanotube.

\subsection*{Acknowledgments}

The authors would like to thank K.~Hashimoto for valuable discussions.
They are also grateful to T.~Fujimori, T.~Nishioka and Y.~Yoshida for
useful comments.
TK is supported by Grand-in-Aid for JSPS Fellows.
The work of MN is supported in part by 
Grant-in Aid for Scientific Research (No. 23740198) 
and by the ``Topological Quantum Phenomena'' 
Grant-in Aid for Scientific Research 
on Innovative Areas (No. 23103515)  
from the Ministry of Education, Culture, Sports, Science and Technology 
(MEXT) of Japan.

\appendix
\section{Instanton construction on the ALE spaces}\label{sec:ADHM}

In this appendix we comment on the ADHM construction for the ALE spaces
\cite{springerlink:10.1007/BF01233429,springerlink:10.1007/BF01444534,Kronheimer:1989zs}.
First let us start with the ADHM method without orbifolding.
We now introduce the ADHM data $(B_1, B_2, I, J)$ satisfying the
following ADHM equations given by
\begin{eqnarray}
 \mu_{\C} & := & \left[ B_1, B_2 \right] + I J = 0 , \\
 \mu_{\R} & := & [B_1, B_1^\dag] + [ B_2, B_2^\dag] + I I^\dag + J^\dag J = 0, 
\end{eqnarray}
where $B_1, B_2 \in \Hom(V, V)$, $I \in \Hom(W, V)$ and $J \in \Hom(V,
W)$.
The dimensions of these vector spaces are dim~$V=k$, dim~$W=N$ for
$k$-instanton configuration of $SU(N)$ gauge theory on $\R^4$.
The instanton moduli space is given by
\begin{equation}
 \M_{N,k} = \left\{ (B_1, B_2, I, J) | \mu_{\C} = \mu_{\R} = 0 \right\}
  / U(k) .
\end{equation}
We have $U(k)$ symmetry for the ADHM data such that 
\begin{equation}
 (B_1, B_2, I, J) \longrightarrow 
  (g B_1 g^{-1}, g B_2 g^{-1}, g I, J g^{-1}), \qquad
  g \in U(k).
\end{equation}

The ALE space is given by resolving the singularity of 
the orbifold $\C/\Gamma$ where $\Gamma$ is a finite subgroup
of $SU(2)$.
We now discuss only 
the case of $\Gamma=A_{n-1}$ for simplicity.
Let $(z_1, z_2)$ be a coordinate of $\C^2$, thus the orbifold $\C^2/\Z_n$
is obtained by identification $(z_1, z_2) \sim (\omega z_1, \bar\omega z_2)$
with $\omega = \exp (2\pi i/n)$.
To study how this identification affects the ADHM equation, we then consider
action of the isometry $(z_1, z_2) \to (e^{\epsilon_1} z_1, e^{\epsilon_2}
z_2)$ on the ADHM data, $(B_1, B_2, I, J) \to (e^{\epsilon_1} B_1,
e^{\epsilon_2} B_2, I, e^{\epsilon_1+\epsilon_2}J)$.
Decomposing the vector spaces as well as the vortex theory
(\ref{decom_vec_sp}) due to the irreducible representation of $\Z_n$,
again these dimensions are dim~$V_m = k^{(m)}$, dim~$W_m = N^{(m)}$.
Then we have
$B_{1,m}\in\Hom(V_m,V_{m+1})$, 
$B_{2,m}\in\Hom(V_m,V_{m-1})$, 
$I_{m}\in\Hom(W_m,V_m)$ and $J_m\in\Hom(V_m,W_m)$.
We can write down the ADHM equation for the ALE space in terms of these
components,
\begin{eqnarray}
 B_{1,m} B_{2,m+1} - B_{2,m} B_{1,m-1} + I_m J_m & = &
  -\zeta_{\C}^{(m)}, \nonumber \\ && \\
   B_{1,m-1} B_{1,m-1}^\dag - B_{1,m}^\dag B_{1,m}
 + B_{2,m} B_{2,m}^\dag - B_{2,m+1}^\dag B_{2,m+1}
 + I_m I_m^\dag - J_m^\dag J_m & = & -\zeta_{\R}^{(m)}, \nonumber \\
\end{eqnarray}
where $\zeta_{\C}^{(m)} \in \C$ and $\zeta_{\R}^{(m)} \in \R$ are
related to blow-up parameters of orbifold singularities, satisfying 
\begin{equation}
 \sum_{m=0}^{n-1} \zeta_{\C}^{(m)} =
 \sum_{m=0}^{n-1} \zeta_{\R}^{(m)} = 0 .
\end{equation}
When all $\zeta_{\C}^{(m)} = \zeta_{\R}^{(m)}=0$, the ALE space goes
back to the singular orbifold.

\begin{figure}[t]
 \begin{center}
  \includegraphics[width=25em]{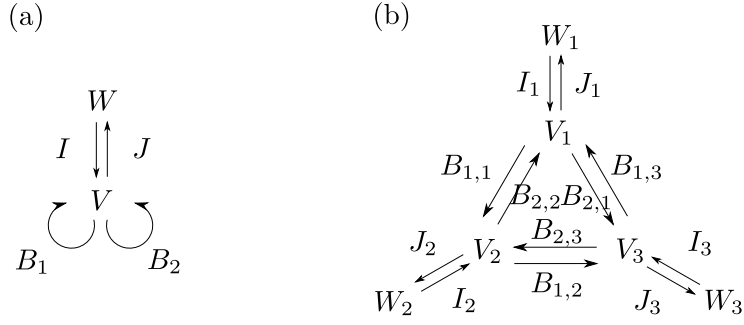}
 \end{center}
 \caption{Quiver diagrams for the moduli spaces of instantons on the
 orbifolds: (a) $\C^2$ and (b) $\C^2/\Z_3$. }
 \label{quiv_fig_inst}
\end{figure}

Fig.~\ref{quiv_fig_inst} shows the quiver diagram for the ADHM data.
Note that it is directly related to Dynkin diagram for the $A_{n-1}$
root system.
If we consider other types of orbifolds, type $D$ or $E$, we have the
corresponding quivers.



\providecommand{\href}[2]{#2}\begingroup\raggedright\endgroup

\end{document}